\documentclass[10pt,twocolumn,floatfix,aps,pre]{revtex4-1}
\usepackage{amsmath,amssymb}
\usepackage{natbib}
\usepackage{graphicx}
\usepackage[T1]{fontenc}
\usepackage[utf8]{inputenc}
\usepackage{times}

\newcommand{\ud}{{\mathrm{d}}}
\newcommand\e{\mathrm{e}}
\newcommand\ii{\mathrm{i}}

\newcommand\vq{\mathbf{q}}
\renewcommand\vr{\mathbf{r}}
\newcommand\vs{\mathbf{s}}
\newcommand\vx{\mathbf{x}}
\newcommand\avg[1]{\left\langle{#1}\right\rangle}

\DeclareMathOperator{\sinc}{sinc}

\DeclareMathOperator{\erf}{erf}

\newcommand\UJF{Universit\'e Joseph Fourier, Grenoble - FRANCE}

\begin{document}
\title{Locating a weak change using diffuse waves (LOCADIFF): 
       theoretical approach and inversion procedure }
\author{Vincent Rossetto}
\email{vincent.rossetto@grenoble.cnrs.fr}
\affiliation{\UJF}
\affiliation{Laboratoire de Physique et Mod\'elisation des Milieux Condens\'es -
  CNRS, Maison des Magist\`eres - BP 166 \\%
  25, avenue des Martyrs - 38042 Grenoble CEDEX - FRANCE}
\author{Ludovic Margerin}
\affiliation{Centre Europ\'een de Recherche et %
  Enseignement en G\'eosciences et Environnement\\
  Universit\'e Aix-Marseille/CNRS - %
  BP 80 13545 Aix-en-Provence, FRANCE}
\author{Thomas Plan\`es}
\author{\'Eric Larose}
\affiliation{Laboratoire de G\'eophysique interne et Tectonophysique - 
             CNRS \& Universit\'e J. Fourier, BP 53 
             38041 Grenoble, FRANCE}

\begin{abstract}
We describe a time-resolved monitoring technique for heterogeneous media. Our
approach is based on the spatial variations of the cross-coherence of coda
waveforms acquired at fixed positions but at different dates. To locate and
characterize a weak change that occurred between successive acquisitions, we
use a maximum likelihood approach combined with a diffusive propagation model.
We illustrate this technique, called LOCADIFF, with numerical simulations.  In
several illustrative examples, we show that the change can be located with a
precision of a few wavelengths and its effective scattering cross-section can
be retrieved. The precision of the method depending on the number of source
receiver pairs, time window in the coda, and errors in the propagation model is
investigated. Limits of applications of the technique to real-world
experiments are discussed. 
\end{abstract}

\maketitle

\section{Introduction}
\label{s:intro}
Elastic and acoustic waves constitute one of the primary tools to detect and
locate temporal changes in natural or man-made structures. If the waves do not
interact with any other obstacle than the target, conventional imaging
techniques are usually based on geometrical considerations. A controlled pulse
emitted into the medium is scattered by the target and the echos are recorded
with a receiver. These techniques can be improved using several sources and
detectors, and extended to locating several targets at the same time. As long
as the typical propagation time in the medium is much smaller than the
scattering mean free time, i.e. the average time between two scattering events,
we are in the single scattering regime. In this case, the resolution for
detecting and locating a change is limited by the Fresnel zone $\sqrt{\lambda
L}$, with $L$ the typical propagation distance in the medium. Applications in
every day life abound: they cover high-stake fields like ultrasonic medical
imaging, non-destructive testing, seismic exploration, radar aircraft location
or sonar. 

This simple picture does not apply in heterogeneous media such as polycrystals,
concrete, or volcanoes. Imaging these materials in a non-destructive way is an
important issue for miscellaneous applications like monitoring, ageing or
damage assessment. In heterogeneous media, ray theory is not relevant because
the scattering mean free time is much smaller than the typical record duration.
A pulse emitted into the medium experiences numerous scattering events and the
output signal recorded at large distance from the source displays complex
details that depend on the interactions between the wave and each of the
scatterers. Beyond a distance called transport mean free path~$\ell^\star$, the
memory of the initial direction of propagation is lost.  In this regime, the
average energy distribution in the medium evolves as a diffusion process and it
is relevant to describe wave propagation using probabilities. 

The problem of locating an isolated change in a multiple scattering sample has
received some attention in the past, particularly in optics. The space and time
correlations of intensity in a speckle pattern probed by one or more receivers
allow one to observe the diffusion of scatterers~\cite{pine1988,berkovits1991}.
On one hand, diffusive wave spectroscopy~\cite{cowan2002} and its variants have
become standard tools for investigating collective changes in the medium. On
the other hand previous authors~\cite{nieuwenhuizen1993} have shown that a
local change (the perturbation) within a collection of scatterers (the
background) essentially acts as a dipole source of intensity. Intensity
variations enable the detection and location of a crack from observations in
transmission~\cite{feng1991,vanneste1993}, or more generally to locate an
object with known characteristics~\cite{den_outer1993,vanrossum1999}. The weak
sensitivity of the method has been illustrated by numerical studies
\cite{vanneste1993}. Indeed, a large amount of ensemble or frequency averaging
(typically 100 realizations) is required to distinguish the intensity
fluctuation caused by the defect from the background speckle pattern. From a
theoretical point of view, the weak sensitivity can be traced back to the
cancellation of diagrams that dominate the waveform decorrelation, a
cancellation which is imposed by the optical theorem. This renders techniques
based on intensity variations almost inapplicable to solid media. These points
will be further illustrated below.

In acoustics, one can commonly record a large number of signals with perfect
temporal and spatial resolution, which is advantageous compared to optics. A
pulse emitted into a medium gives rise to long time records with a pronounced
coda, a term which refers to the arrivals following the ballistic pulse.
Several techniques use the coda to retrieve information on the evolution of the
medium. In seismology, the monitoring of temporal changes in the crust was
initiated in the mid-80's, using repeating small earthquakes on faults
\cite{poupinet1984}. Later on the method was applied to volcanoes and revealed
temporal changes of velocity prior to eruptions \cite{ratdomopurbo1995}. The
method was transposed to the laboratory and popularized under the terms diffuse
acoustic wave spectroscopy (DAWS)~\cite{cowan2000}, or coda-wave interferometry
(CWI)~\cite{snieder2002,pacheco2005}. In these approaches, changes of waveforms
in the coda are interpreted in terms of travel time variations, a technique
that is very sensitive~\cite{larose2008} to detect weak changes, but gives
little information concerning the location of the change. To first order,
global velocity changes in the medium result in a stretching of the waveforms
\cite{ratdomopurbo1995,lobkis2003,larose2008,brenguier2008} but the
interpretation of a local change in terms of travel time fluctuation remains
problematic. Recently DAWS has been used in damage
monitoring~\cite{michaels2005,tremblay2010} but a large range of other
applications are possible \cite{snieder2007}. For a broad review of
applications of CWI in geophysics, we refer to \cite{poupinet2008}. Also based
on the concept of correlation, techniques have been developed to recover the
Green's function in an open medium based on the cross correlation of noise
signals~\cite{derode2003,larose2006b,wapenaar2005,wapenaar2006,gouedard2008},
These noise-based Green's functions can in turn be used in a passive image
interferometry technique~ with applications in volcanology and fault monitoring
\cite{sens-schoenfelder2006,wegler2007,brenguier2008}. Recently, Aubry and
Derode \cite{aubry2009} proposed an alternative technique based on the singular
value decomposition of the propagator, but this technique is limited to a
sufficiently strong extra scatterer and is not sensitive to weak perturbations. 

In this article, we report on a different approach to locate a small isolated
change. Our LOCADIFF technique uses the correlations between time windows in
the late coda for several pairs of sources and receivers. A numerical model of
the medium is then used to compute the most likely position of the weak change,
in terms of probability. We start our description of the work by observing the
correlation loss of signals induced by the weak change in a finite difference
numerical simulation (Section \ref{s:observation}).  Using the theory of
multiple scattering~\cite{vanrossum1999}, we derive an expression of the
decorrelation induced by a weak change in Section~\ref{s:theory}. We then
present the inversion technique, based on the maximum likelihood principle in
Section~\ref{s:inversion}. We discuss the accuracy of the technique and
possible improvements in Section~\ref{s:discussion}.

\section{Observations of correlation loss after a weak change}
\label{s:observation}
It is already known that a weak change can be detected in a scattering medium
because it slightly modifies the coda of the Green's functions. The amount of
modification is usually quantified by measuring the cross-correlation between
waveforms recorded at different times~\cite{poupinet1984}. We illustrate the
signal processing with the aid of finite-difference simulations of the wave
equation in a medium containing a large number of identical scatterers.
\subsection{Numerical simulations of wave propagation} 
\begin{figure}[!htbp]
 \centering
 \includegraphics[width=8cm]{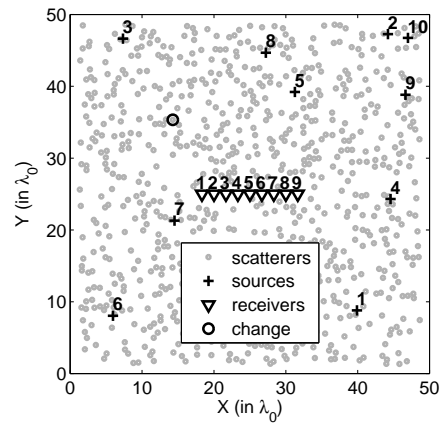}
 \caption{Distribution of sources, receivers, scatterers in the
 numerical simulation. The scatterer to remove is surrounded in gray.}
 \label{f:simu2d}
\end{figure}

As a first
investigation, we perform 2D numerical experiments of acoustic wave propagation
in heterogeneous open media ~\footnote{The code named ACEL has been developed
by M. Tanter, Institut Langevin, Paris France. More details in
\texttt{http://www.institut-langevin.espci.fr/Mickael-Tanter,143}}.  Using a
finite difference centered scheme, we solve the wave equation with absorbing
boundary conditions; the dimension of the simulation grid is $50\lambda_0
\times 50 \lambda_0$, with a spatial discretization step $\lambda_0/30$, where
$\lambda_0$ is the central wavelength. Synthetic data are computed on a linear
array of 9 receivers located at the center of the medium and 10 sources are
randomly distributed over the grid. Sources and receivers are kept fixed
throughout the experiments (see Fig.~\ref{f:simu2d}).  To mimic a multiple
scattering medium, 800 empty cavities of diameter $\lambda_0/3$ are randomly
distributed over the grid. In the frequency band of interest the average
scattering cross-section was numerically estimated as $\Sigma=1.6\lambda_0$,
along with the transport cross section $\Sigma^\star=1.1\lambda_0$.
Table~\ref{t:simu2d} summarizes the physical properties of the simulated
medium, including the number of scatterers (with density $n$), the transport
mean free path $\ell^\star=\frac{1}{n\Sigma^{\star}}$, the diffusion constant
$D=\frac{c\ell^{\star}}{2}$ and the Thouless time $\tau_D=\frac{R^2}{6D}$,
where $R^2$ is the mean squared distance between sources and receivers. Note
that these quantities are evaluated under the ``independent scattering
approximation'', which assumes that the waves never visit the same scatterer
twice.

\begin{figure}[!htbp]
 \centering
 \includegraphics[width=8cm]{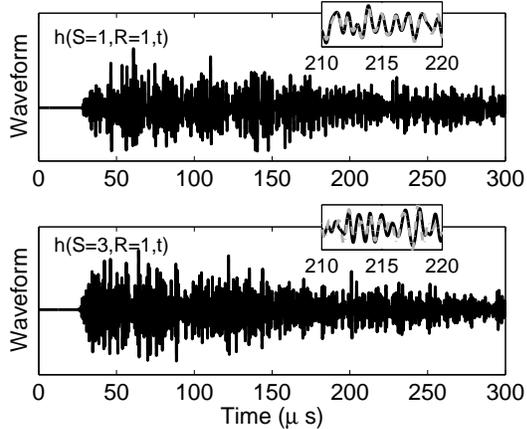}
 \caption{\label{f:waveforms2d}
 Normalized waveforms $h(t)$ obtained at receiver 1, 
 for source number 1 (top) and 3
 (bottom). Inset: zoom into the coda. 
 The black solid (resp gray broken) line
 corresponds to the record acquired before (resp. after) the change.
 For illustration purpose, the value of~$T_0$ was set 
 to~$1\,\mathrm{\mu s}$. }
\end{figure}

\begin{table}[!htbp]
 \begin{tabular}{|c|c|c|c|}
 \hline
 Parameter & notation & value \\
 \hline \hline
 Number of scatterers & & 800\\
 \hline
 Transport mean free path &$\ell^\star$ &$2.8 \lambda_0$ \\
 \hline
 &$k\ell^{\star}$ & 18 \\
 \hline
 Diffusion constant &$D$ & $1.4\,\lambda_0^2/T_0$ \\
 \hline
 Thouless time & $\tau_D$ & $68\, T_0$ \\
 \hline
 Coda decay time (leakage) & $\tau_\sigma$ & $240\, T_0$ \\
 \hline
 \end{tabular}
 \caption{\label{t:simu2d}
 Physical parameters of the simulations in normalized units.}
\end{table}
The signal $e(t)$ emitted by each source is a pulse with central frequency
$f_0$ and a Gaussian envelope (100\% bandwitdh at -6dB).  Using source~$i$ we
record with receiver~$j$ the signal~$h_{ij}(t)$ during~$300$ oscillations of
period~$T_0$. Typical waveforms $h_{ij}(t)$ are plotted in
Fig.~\ref{f:waveforms2d}. The long tail of the record in
Fig.~\ref{f:waveforms2d} corresponds to arrival of partial waves that have been
scattered several times. Notice that the ballistic arrival is not
distinguishable in the waveforms of Figure~\ref{f:waveforms2d}. Long coda and
lack of ballistic arrival constitute evidences that we are in a strongly
scattering regime, in agreement with our estimates of the transport mean free
path. During the first run of the simulation, $10\times 9$ impulse responses
$h_{ij}$ are recorded and stored. On a second run, one scatterer is removed and
another set of impulse responses $h'_{ij}$ is evaluated. Both $h_{ij}(t)$ and
$h'_{ij}(t)$ display long codas lasting a large number of ballistic times. 

\subsection{Detection of an isolated change}
The details of the complex waveforms shown in Figure \ref{f:waveforms2d} are
highly sensitive to the positions of the the scatterers.  Each waveform can be
understood as a fingerprint of the medium. As our goal is to detect a single
scatterer's change, we need to exploit the information contained in both the
amplitude and phase of the signals. A comparison of the records $h_{ij}(t)$ and
$h'_{ij}(t)$ reveals that for short coda times up to $100\,T_0 \approx 30
\tau^*$, no difference is visible in the signals. We observe small differences
between the waveforms at later times that are solely due to the change in the
medium. Figure~\ref{f:waveforms2d} (bottom) shows an example of such
differences for $h_{3,1}(t)$ and $h'_{3,1}(t)$. The observed decorrelation is
too large to be attributed to numerical noise, there is thus evidence that the
coda waveforms are sensitive to the removal of only one scatterer.

The differences between the waveforms $h_{ij}$ and $h'_{ij}$ are quantified by
the decorrelation, or correlation loss, between $h_{ij}$ and $h'_{ij}$. The
decorrelation is computed in a time window of duration $2T$ centered on $t$
using the formula: 
\begin{equation}
 K_{ij}(t)=1-\frac{\int_{t-T}^{t+T} h_{ij}(u) \,h'_{ij}(u) \ud u}
 {\sqrt{\int_{t-T}^{t+T} h_{ij}(u)^2 \ud u
 \int_{t-T}^{t+T} h'_{ij}(u')^2 \ud{u'}}}.
 \label{e:decorrelation}
\end{equation}
The typical width of the time window $T$ is of the order of $5T_0$.
Experimentally, enlarging~$T$ partly eliminates the effect of noise and reduces
the fluctuations of the correlation coefficient. Nonetheless, using a large
value for~$T$ results in considering simultaneously paths with very different
lengths. We address this important point in Section~\ref{s:decorrelation}.

\subsection{Spatial dependence of the decorrelation}
In Figure~\ref{f:waveforms2d}, it is noticeable that the differences between
$h_{1,1}(t)$ and $h'_{1,1}(t)$ (top), are much smaller than the differences
between $h_{3,1}(t)$ and $h'_{3,1}(t)$ (bottom), even in the late coda. The
decorrelations computed over the interval $[210\,T_0,\,220\,T_0]$ are
$K_{1,1}(215\,T_0)=5\%$ and $K_{3,1}(215\,T_0)=27\%$, respectively.
Consequently, the amount of decorrelation depends on the positions of the
source and receiver with respect to the local change, a property which holds
even in very late time windows in the coda. For a given configuration of
source-receiver pairs, we obtain a set of observed decorrelations, which are
characteristic of the relative locations of the source, receiver and change in
the multiple scattering medium. We will now demonstrate the possibility to
locate the change and estimate its cross-section from the knowledge of the
source and receiver positions and the corresponding decorrelation coefficients.
To do so, we develop a theoretical model to predict the decorrelation
coefficient of waves induced by the addition of a change in a heterogeneous
medium, in the diffusive regime. We recall in the next section the necessary
elements from multiple scattering theory.  

\section{Wave scattering theory}
\label{s:theory}
We assume that the medium can be represented as a matrix with embedded
inclusions. Only the scalar case is considered here. The scattering properties
of an inclusion will be described by its $\mathcal{T}$ matrix, defined in
operator notation as \cite{Sheng.2006,Economou.2006}: 
\begin{equation}
 G_1 = G_0 + G_0\mathcal{T}G_0
 \label{def.t}
\end{equation}
where $G_0$ is the retarded free space Green function and $G_1$ is the Green
function in the presence of the scatterer. For a non absorbing scatterer,
energy conservation implies the following optical theorem: 
\begin{equation}
 -\frac{\Im\mathcal{T}(\omega)}{k_0}=\sigma(\omega),
 \label{optical.theorem}
\end{equation}
where $\sigma$ is the scatterer cross-section.

\subsection{Correlations between two slightly different media}
\label{s:correlations}
\begin{figure}
 \centering
 \includegraphics[width=5cm]{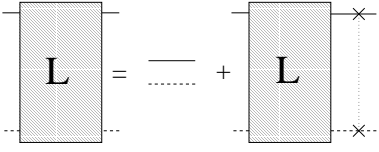}
 \caption{\label{f:diagL} The Bethe-Salpeter equation defining the
so-called ladder operator $L$. The solid line and dashed line represent the
retarded ensemble averaged Green's function and its complex conjugate,
respectively. The dotted line connecting the two vertices indicates
that they represent the same scatterer.}
\end{figure}

We want to predict the decorrelation of waveforms in a medium where
a small change occurs. Although we will employ a statistical approach based on
ensemble averages, in general we have access to only one realization of
the random process. Therefore we introduce the following estimator of the
cross-correlation function based on the observation of a single coda: 
\begin{equation}
 \Gamma(t,\,\tau)=\frac1{2T}
 \int_{t-T}^{t+T} \psi^2(t'+\tau/2)\psi^1(t'-\tau/2) \ud{t'},
 \label{def.Gamma}
\end{equation}
where $\psi$ is the scalar field. The superscript $2$ refers to the medium in
presence of an extra defect while the superscript~$1$ refers to the medium
without it.  We have introduced an analog of the Wigner function which is most
convenient to analyze non-stationary signals. The empirical cross-correlation
can be decomposed over internal and external frequencies $\omega$ and $\Omega$,
respectively: 
\begin{equation}
 \Gamma(t,\,\tau)=\frac1{(2\pi)^2}
 \int_{-\infty}^{\infty} \ud\Omega \int_{-\infty}^{\infty} \ud\omega
 \Tilde\Gamma(\Omega,\,\omega)\,\exp[-\ii(\Omega t+\omega\tau)]
 \label{Gamma.freq}
\end{equation}
where the frequency-domain cross-correlation reads:
\begin{equation}
 \Tilde\Gamma(\Omega,\,\omega)= \sinc(\Omega T) \,
 \psi^2(\omega+\Omega/2)
 \psi^1(\omega-\Omega/2)^*.
 \label{TF.Gamma}
\end{equation}
We see that the width of the time window, $2T$, has a minor effect only. 
Equation \eqref{TF.Gamma} shows that we have to compute the quantity
\begin{equation}
 \avg{G^2(\omega+\Omega/2)G^1(\omega-\Omega/2)^*}.
 \label{def.corr}
\end{equation}
In equation \eqref{def.corr}, $G$ is the retarded Green's function.  We will
denote by $\mathcal{T}_0$ the $\mathcal T$-matrix of the additional defect
which is assumed to appear at the position $\vx_0$.  In diagrammatic notations,
such as the one employed in Figure~\ref{f:diagL} , the $\mathcal{T}$ matrices
are represented by crosses. The transport of energy in the scattering medium is
described by the ladder operator~$L$, which is defined by the diagrammatic
self-consistent equation shown in Figure~\ref{f:diagL}
\cite{Sheng.2006,akkermans2007}.

We use the field-field correlation function in the coda
\begin{multline}
 \Tilde\Gamma(\omega,\,\Omega,\,\vs,\,\vx_0,\,\vr)=
 \int\ud{\vr_1}\int\ud{\vr_2}\, 
 \Tilde P_0(\omega,\,\Omega,\,\vs,\,\vr_1)\\
 \times \Tilde L_e(\omega,\,\Omega,\,\vr_1,\,\vx_0,\,\vr_2)
 \Tilde P_0(\omega,\,\Omega,\,\vr_2,\,\vr).
 \label{field.corr}
\end{multline}
Quantities labelled with~\~{} are implicitely evaluated at inner frequency
$\omega$ and outer frequency $\Omega$. The ladder propagator $L_e$ describes
the transport of correlations in a sequence of scattering events in the medium
with an extra scatterer. $\Tilde P_0(\vs,\,\vr_1)$ and $\Tilde
P_0(\vr_2,\,\vr)$ describe the ballistic propagation from the source to the
first scattering event, and from the last scattering event to the detector,
respectively: 
\begin{equation}
 \Tilde P_0(\vr_1,\,\vr_2)=
 \frac{\e^{-R/\ell}}{(4\pi R)^2}\,\e^{\ii\Omega R/c}
 \label{ex.P0}
\end{equation}
where $R = |\vr_2 - \vr_1|$ and $c = \partial_\omega k_0(\omega)$ is the group
velocity at the frequency $\omega$. The ladder propagator with the
extra-scatterer $L_e$ is related to the ladder propagator without the
extra-scatterer $L$ as follows \cite{nieuwenhuizen1993}:
\begin{figure}[!htbp]
 \centering
 \includegraphics[width=6cm]{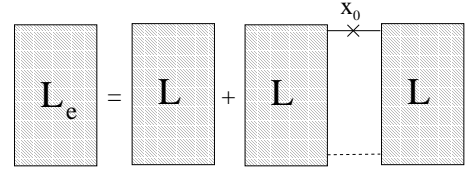}
 \caption{\label{f:diagLe} The diagram of the ladder operator
with an extra scatterer. The extra scatterer is sandwiched between two ladder
operators. Note that we have neglected the possibility that ensemble averaged
Green's functions connect the extra scatterer with the source and/or the
receiver which assumes that it is located at least one mean free path away
from all sources and receivers.}
\end{figure}

\begin{multline}
 L_e(\vs,\,\vx_0,\,\vr)
 =L(\vs,\,\vr)+ \\ \int \!\! \ud{\vr_1}\int\!\! \ud{\vr_2}
 L(\vs,\,\vr_1)J(\vr_1,\,\vx_0,\,\vr_2)L(\vr_2,\,\vr).
 \label{def.Le}
\end{multline}
In Equation~\eqref{def.Le}, also represented by the diagram depicted in
Figure~\ref{f:diagLe}, the first term represents the scattering paths that do
not see the defect, while the second term describes the paths that visit the
defect once. As we are in a regime of weak interaction between the field and
the scatterer higher-order terms can be neglected. We define the operator $J$
that connects the two ladders by
\begin{multline}
 J(\vr,\,\vx_0,\,\vr')=\\
 \int \!\! \ud{\vr_1}\int \!\! \ud{\vr_2}\,
 G(\vr,\,\vr_1)\mathcal{T}(\vr_1,\,\vr_2)G(\vr_2,\,\vr')G(\vr,\,\vr')^*,
 \label{def.J}
\end{multline} 
where $G$ denotes the ensemble averaged Green's function. In the mesoscopic
regime,~$J$ is evaluated to lowest order of the small quantity $1/(k_0\ell)
\ll1$ for a point scatterer:
\begin{equation}
 J(\vr,\,\vx_0,\,\vr')\simeq
 -\frac{\ii \ell^2\mathcal{T}_0}{8\pi k_0}\delta^{(3)}(\vr-\vx_0)
 \delta^{(3)}(\vx_0-\vr').
 \label{approx.J}
\end{equation}
Inserting expression \eqref{approx.J} into equation \eqref{field.corr} one
obtains:
\begin{multline}
 \Tilde\Gamma(\vs,\,\vx_0,\,\vr)=
 \int \ud{\vr_1} \int\ud{\vr_2}
 \Tilde P_0(\vs,\,\vr_1)L(\vr_1,\,\vr_2) \Tilde P_0(\vr_2,\,\vr)\\
 -\int \ud{\vr_1} \int \ud{\vr_2} 
 \Tilde P_0(\vs,\,\vr_1)\Tilde L(\vr_1,\,\vx_0)
 \frac{\ii\ell^2\mathcal{T}_0}{8\pi k}
 \Tilde L(\vx_0,\,\vr_2)\Tilde P_0(\vr_2,\,\vr),
 \label{Gamma.defect}
\end{multline}
where the first term is the diffuse intensity in the medium without extra
scatterer and the second integral is an interference term caused by the
extra scatterer. In the slowly-varying envelope approximation, the integrals
can be evaluated to give:
\begin{equation}
 \Tilde \Gamma(\vs,\,\vx_0,\,\vr)=
 \frac{\ell^2}{4\pi^2} \Tilde L(\vs,\,\vr)
 -\frac{\ell^2}{4\pi^2} \Tilde L(\vs,\,\vx_0)
 \frac{\ii\ell^2\mathcal{T}_0}{8\pi k_0} \Tilde L(\vx_0,\,\vr).
 \label{Gamma.approx}
\end{equation}
In the diffusive regime, the propagator of the wave intensity in the medium
filled with scatterers, $\Tilde P_d$, is the solution of the following
equation
\begin{equation}
 \left(-\ii\Omega-D\nabla^2_{\vr_2}\right)\Tilde P_d(\vr_1,\,\vr_2)=
 \delta^{(3)}(\vr_1-\vr_2),
 \label{e:diffusion}
\end{equation}
where~$D$ is the diffusivity. The ladder $L$ is related to $\Tilde P_d$ by 
$ \Tilde L(\vr_1,\,\vr_2)=\frac{4\pi c}{\ell^2} \Tilde P_d(\vr_1,\,\vr_2).$
Using these notation the diffuse intensity for a unit point source satisfies:
\begin{equation}
 \Tilde\Gamma(\vs,\,\vx_0,\,\vr)=
 \frac{c}{4\pi} \Tilde P_d(\vs,\,\vr) 
 +\frac{c}{4\pi} \Tilde P_d(\vs,\,\vx_0) \frac{\ii c \mathcal{T}_0}{2k_0}
 \Tilde P_d(\vx_0,\,\vr).
 \label{Gamma.approx2}
\end{equation}
In order to obtain the correlation function in the time domain, we double
invert the Fourier transform over the variables $\omega$ and $\Omega$. We
further assume that the signal has been filtered in a narrow frequency band
$\Delta\omega$ in which the scattering properties vary little. Upon integration
over $\omega$ and application of the optical theorem \eqref{optical.theorem},
the correlation function for a unit point-source normalized by the bandwidth
$\Delta\omega$ reads:
\begin{multline}
 \Gamma(\vs,\,\vx_0,\,\vr,\,t)= P_d(\vs,\,\vr,\,t)\\
 -\frac{c\sigma}{2}\int_0^t\ud u\, P_d(\vs,\,\vx_0,\,u)
 P_d(\vx_0,\,\vr,\,t-u).
 \label{proba.defect}
\end{multline}
We have therefore obtained the theoretical decorrelation
$K(\vx_0,t)=\frac{c\sigma}{2}Q(\vs,\,\vx_0,\,\vr,\,t)$, where
\begin{equation}
 Q(\vs,\,\vx_0,\,\vr,\,t) =
 \frac{\int_0^t\ud u\, P_d(\vs,\,\vx_0,\,u)P_d(\vx_0,\,\vr,\,t-u)}
 {P_d(\vs,\,\vr,\,t)}.
 \label{def.Q}
\end{equation}
The negative sign in~\eqref{proba.defect} comes from the optical theorem
(energy conservation) and ensures that the cross-coherence is less than one.
The derivation presented in this section does not depend on the form of
Equation~\eqref{e:diffusion}, which means that solutions to a more accurate
transport equation can be substituted to~$P_d$. Note that for a resonant point
scatterer, $\sigma$ can be substituted with $\lambda_0^2/\pi$.

\subsection{Computation of the decorrelation formula}
\label{s:decorrelation}
We observe that the decorrelation~\eqref{def.Q} can be computed if the
function~$P_d$ is known. In the general case where the diffusivity~$D$ depends
on the position, the function $P_d$ can only be numerically estimated, provided
that the spatial dependence of~$D$ is known. In practice, the decorrelation
coefficient can be reasonably rapidly computed if one assumes that the value
of~$D$ is approximately uniform in the medium. We investigate the amount of
variation for~$D$ in Section~\ref{s:diffusivity}.

If the medium is absorbing, the same issue arises. In media with a uniform
absorption time $\kappa^{-1}$, the absorption affects the numerator of~$Q$ in
\eqref{def.Q} by a factor $\exp[-\kappa u -\kappa (t-u)] =\exp[-\kappa t]$ and
the denominator by a factor~$\exp[-\kappa t]$. Therefore, uniform absorption
effects cancel out in the normalized decorrelation function, which is a genuine
advantage of the present technique. In the case where absorption is
non-uniform, it will affect differently $P_d(\vs,\,\vx_0)$ and
$P_d(\vx_0,\,\vr)$ and the observed decorrelation pattern may be partly
ascribed to the spatial variations of absorption. Consider a medium with
constant diffusivity~$D$ and absorption~$\kappa$. The solution of the diffusion
equation~\eqref{e:diffusion} in an infinite $d$-dimensional medium is 
\begin{equation}
 P_d(\vr_1,\,\vr_2,\,t)=\frac{1} {(4\pi Dt)^{d/2}}
 \exp\left[-\kappa t-\frac{(\vr_2-\vr_1)^2}{4Dt}\right].
 \label{e:diffusiond}
\end{equation}
In the case of a 3-D infinite medium, a usual Laplace transform calculation
gives the exact result:
\begin{equation}
 Q(\vs,\,\vx,\,\vr,\,t)=\frac{1}{4\pi D}
 \left( \frac{1}{s}+\frac{1}{r}\right) 
 \exp\left[\frac{R^2-(s+r)^2}{4Dt}\right].
 \label{formula.diffusion}
\end{equation}
where we have introduce the notations $s=\|\vs-\vx\|$, $r=\|\vr-\vx\|$ and
$R=\|\vs-\vr\|$.  We observe that $Q$ is a function with elliptic contour lines
multiplied by simple poles located at~$\vs$ and $\vr$. Of course, if~$\vr=\vs$,
we recover the formula derived in Ref.~\cite{snieder2002} for an infinite
medium. This formula is generally not applicable under this form because the
transducers are usually located at the surface of the system. However, if the
boundary conditions are sufficiently simple, the formula
(\ref{formula.diffusion}) can be used as a building block to derive more
complicated solutions, as shown in Section~\ref{s:boundaries}. 

\begin{figure}
 \centering
 \includegraphics[width=0.7\linewidth]{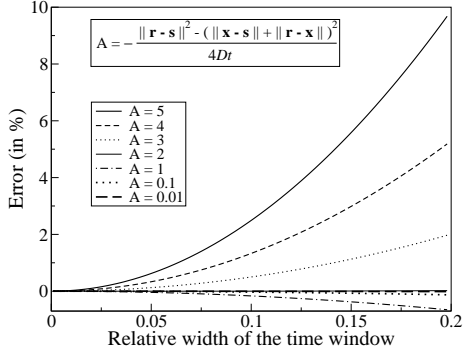}
 \caption{\label{f:error} Deviation of the average of~$Q$ on the
 time interval~$[t-T,\,t+T]$ with repect to~$Q(t)$. The correction grows
 more rapidly for large values of the argument in the exponential 
 of~\eqref{formula.diffusion}, denoted by A in this Figure.}
\end{figure}

In formula~\eqref{proba.defect} we neglect two constraints. First, we assume
that the change occurs at a minimum distance of the order of one mean free path
from the source and the receiver. Second, we neglect the finite velocity of
the wave, in other words, the contribution for times $u,t-u< R/c$ in the
integral~\eqref{proba.defect} should be removed. The contribution of short
times $u < R/c $ in~\eqref{proba.defect} is negligible as soon as $ct \gg R >
\ell^\star$. The computation of the decorrelation coefficients~$K_{ij}(t)$
must be done with~$T$ larger than a few oscillation periods of the wave. Using
formula~\eqref{formula.diffusion}, we can estimate the correction due to this
averaging as a function of~$T/t$. To do so, we compute the average
of~\eqref{formula.diffusion} on the interval $[t-T,\,t+T]$ and divide by the
value of~$Q$ at~$t$. We obtain a curve of relative correction as a function
of~$T/t$ which is independent of any other parameters and which is displayed on
Figure~\ref{f:error}. In most applications, the correction will be typically
less than 10\%. 

\subsection{Intensity variations vs field correlations}
As recalled in the introduction, a number of investigations on the monitoring
of complex media have focused on the detection of intensity variations induced
by local changes of the scattering properties. We will show that in the
diffusive regime, intensity variations are much less sensitive to local changes
than field correlations. To do so, we calculate the perturbation of the ladder
propagator induced by an extra-scatterer following the approach developed in
reference \cite{nieuwenhuizen1993}. In addition to the diagram depicted in
Figure \ref{f:diagLe}, two other diagrams contribute to intensity variations:
1) a diagram with a single cross on the lower line and 2) a diagram with one
cross on each line which are connected by dashed line. In the diffusive regime
and for a non-absorbing defect, we obtain the intensity perturbation to
lowest-order in $\vq$ and $\vq'$ in the form: 
\begin{multline}
\delta L_e^{I}(\vs,\,\vx,\,\vr;\,t) = \frac{\ell^4 \mathcal{T} \mathcal{T}^\star}{48 \pi^2}
 \int^{+\infty}_{-\infty} \frac{\ud\Omega}{2 \pi} 
 \iiint_{\mathbb{R}^3 \times \mathbb{R}^3} \ud^3\vq\,\ud^3\vq' \\ 
 L(\vq;\Omega) \e^{\ii\vq\cdot(\vr-\vx)} (\vq\cdot\vq') L(\vq';\,\Omega) 
 \e^{\ii\vq'\cdot(\vx-\vs)} \e^{-\ii \Omega t}. 
\end{multline}
In the Fourier domain, the ladder propagator in the diffusion regime writes:
\begin{equation}
 L(\vq;\,\Omega) = \dfrac{4 \pi}{(2\pi)^3 \ell^2 \left(q^2 \ell/3 -\ii \Omega/c \right) }.
\end{equation}
After integration over the wavenumbers $\vq$, $\vq'$ and the frequency $\Omega$, we obtain:
\begin{multline}
 \delta L_e^{I}(\vs,\,\vx,\,\vr;\,t) = 
 \dfrac{\sigma c}{6 \pi^{1/2} D^{3/2} t^{3/2} } \e^{-R^2/4Dt} \times \\
 \nabla_\vs\cdot\;\left(\nabla_\vr Q(\vs,\,\vx,\,\vr) \right).
\end{multline} 
After calculation of the partial derivatives, we obtain the following formula
for the ladder perturbation induced by an extra scatterer:
\begin{multline}
 \delta L_e^{I}(\vs,\,\vx,\,\vr;t) = 
\dfrac{\sigma c^2 (\vr-\vx) \cdot (\vs-\vx) }
{48 \pi^{3/2} D^{7/2} t^{5/2} r^2 s^2 } \times \\ 
\left(\dfrac{r^3 +s^3}{r s} + \dfrac{(r + s)^3}{2 D t} \right)
 e^{-(r + s)^2/4Dt} 
\end{multline}
The intensity variation exhibits a characteristic pattern with positive and
negative lobes, depending on the cosine of the angle between the source and
receiver as seen from the additional scatterer. Even more important is the
temporal dependence $t^{-5/2}$ which is faster than the temporal decay of the
ladder propagator between the source and receiver. As a consequence, the
sensitivity to the local change decays like $1/t$ in the coda in sharp contrat
to the field correlation which goes to a constant at large lapse time. This
property justifies the popular use of field correlation functions to monitor
temporal changes in evolving media.

\section{The inversion technique}
\subsection{Maximum likelihood of the position}
\label{s:inversion}
In Section~\ref{s:theory}, we have obtained an expression for the expected
decorrelation as a function of the position of the change. The principle of the
inversion technique is to compare a numerical model to experimental data.  The
change is found at the position where numerical and experimental decorrelation
match best. The mismatch is measured by a standard least-squares cost function
($\chi^2$). The inversion technique consists in finding the position~$\vx$ and
the cross-section~$\sigma$ minimizing the function~$\chi^2$.  Such a technique
is also often called a maximum likelihood method.  Let us chose a set of
sources $\vs_i$ ($1\leq i\leq n_s$) and a set of receivers $\vr_j$ ($1\leq
j\leq n_r$), and call $N$ the number of source-receiver pairs (in this case,
$N=n_rn_s$). There is no restriction on their positions, and in particular
source and receiver can be located at the same position. We describe the
technique at fixed time $t$ in the coda.

\begin{figure}[!hbtp]
 \centering
 \includegraphics[width=8cm]{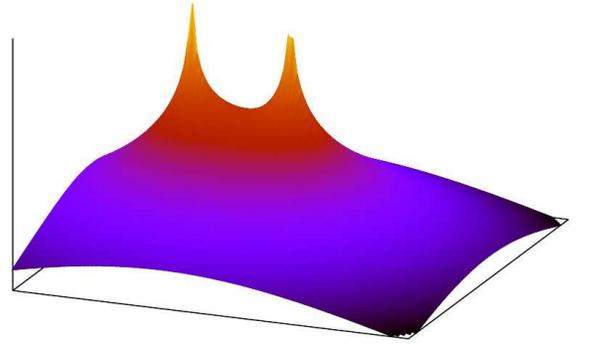}
 \caption{The function $Q_{ij}(\vx)$ in a infinite medium in dimension 3
 with constant diffusivity and absorption computed with formula
\eqref{formula.diffusion}.
 The function is plotted in the plan containing the source, the receiver 
and the change. The value along the $z$ axis (logarithmic) is the sensitivity
to a change at the position in~$(x,y)$. 
 The two peaks correspond to the positions of the source and the receiver.
 The $z$-scale is logarithmic and arbitrary.
 \label{f:Qij}}
\end{figure}
The most restrictive assumption of our approach is that a single defect
affects the experimental values of the decorrelation. The LOCADIFF inversion
technique consists in retrieving the most likely position of this defect by
introducing the cost function:
\begin{equation}
 e(\vx)=\sum_{i,j} \left(K^m_{ij}(t)\right. - K_{ij}(\vx,\,t))^2 /\epsilon^2,
 \label{def.e_exp}
\end{equation}
where $K^m_{ij}(t)$ denotes the experimental measurements of the
decorrelation and the coefficients~$K_{ij}(\vx,\,t)$ are the theoretical
decorrelations assuming that the defect is located at $\vx$. The typical
fluctuations on the measured decorrelations are encapsulated in the parameter
$\epsilon$.

To find the value of the scattering cross-section~$\sigma$, also unknown, we
remark that $e(\vx)$ is, as a function of~$\sigma$, a polynomial of degree two.
There is therefore a minimum depending on $\vx$ at
\begin{equation}
 \sigma_{\text{opt}}(\vx)=\frac 2c 
 \frac{\sum_{i,j} K^m_{ij}(t)Q_{ij}(\vx,\,t)}{\sum_{i,j} Q_{ij}(\vx,\,t)^2}.
 \label{e:sigmaopt}
\end{equation}
We reintroduce the value of~$\sigma_{\text{opt}}$ into the
expression~\eqref{def.e_exp} and get the optimized error function
\begin{equation}
 e_{\text{opt}}(\vx)=\sum_{i,j} \dfrac{K_{ij}(t)^2}{\epsilon^2} -
 \frac{\left(\sum_{i,j} K_{ij}(t)Q_{ij}(\vx,\,t)\right)^2}
 {\epsilon^2 \sum_{i,j}Q_{ij}(\vx,\,t)^2} 
 \label{e:eopt}
\end{equation}
which does not depend on~$\sigma$ anymore. The most likely position of the
defect is the position~$\vx_0$ of the minimum of~$e_{\text{opt}}$. The value of
the cross-section is $\sigma_{\text{opt}}(\vx_0)$ obtained from
Equation~\eqref{e:sigmaopt}.

To give an interpretation to the values of $e(\vx)$, it is customary to
normalize it in the following way
\begin{equation} 
 \chi^2_n(\vx)= \frac{e(\vx)}{f}
 \label{def.chi2}
\end{equation}
where $f = N - 4$ is the number of degrees of freedom, since four model
parameters -the cross-section and the cartesian coordinates of the defect- are
to be estimated. The quantity $\chi^2_n(\vx)$ has the following
interpretations. If $\chi^2_n(\vx)\gg1$, it is very unlikely that the
point~$\vx_0$ is actually located at~$\vx$. If $\chi^2_n(\vx)\simeq1$ the
point~$\vx$ is a good candidate for~$\vx_0$. If $\chi^2_n(\vx_0)\ll1$, there
is a large area where~$\chi^2_n(\vx)<1$ which means that the inversion could
not locate precisely the change because the value of $\epsilon$ is too large.
In other words the quality of measurements is too poor to give any satisfactory
result. It is possible to use $\chi^2_n(\vx)$ to obtain the probability 
density that the defect has appeared at the point $\vx$, which we define as:
\begin{equation}
 p(\vx)=\frac1C\exp\left[-\frac12N\chi^2_n(\vx)\right] 
 =\frac1C\exp\left[-\frac{e(\vx)}{2\epsilon^2}\right]
 \label{chi2.proba}
\end{equation}
where $C$ is a normalization constant such that $\int p(\vx)\ud\vx=1$ (see the
appendix for a derivation of this formula).

\begin{figure}
 \includegraphics[width=\linewidth]{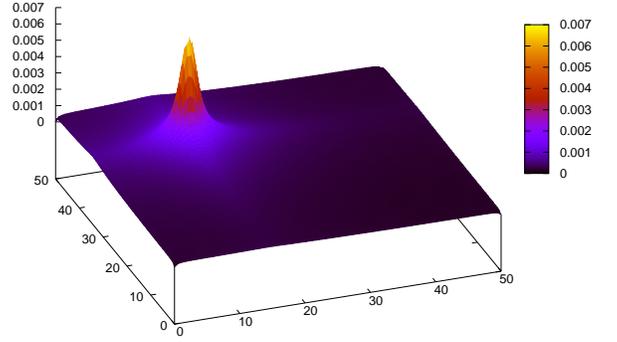}
 \caption{\label{f:probability} Density of probability
 for the position of the moving scatterer.}
\end{figure}

\subsection{Resolution versus number of source-receiver pairs}
\label{s:resolution}
\begin{figure}
 \centering
 \includegraphics[width=0.4\linewidth]{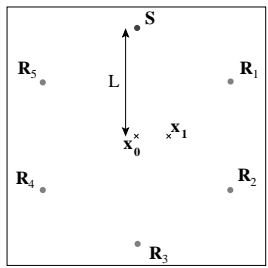}
 \caption{\label{f:cercle} Description of the numerical setup
 used for investigating the accuracy of the inversion technique.
 The example is shown with $N=5$. The other parameters of the numerical
 simulations are : $L=10$, $D=1$, $c=1$. The change~$\vx_0$ is located
 at the center of the circle and is used in Section~\ref{s:resolution}
 to study the optimal spatial resolution of the inversion. The change
 located at point~$\vx_1$ is used to study the robustness of the inversion
 technique against measurement errors on the determination of~$D$ in 
 Section~\ref{s:diffusivity}.}
\end{figure}

To investigate the resolution of the inversion technique depending on the
parameters of the likelihood maximization, we use a numerical approach. We
compute the best achievable resolution regardless of all experimental
difficulties that potentially degrade the accuracy of the location. We use an
ideal set-up made of one source and $N$ receivers regularly distributed on a
circle (see figure~\ref{f:cercle}).  We introduce a change at the center of the
circle by adding a single scatterer with cross-section $\sigma$. For each pair
of receiver, we compute synthetic data through application of the formula
(\ref{formula.diffusion}). The Thouless time~$\tau_D$ is defined as $L^2/D$. As
a measure of the precision, a resolution length $\delta$ is introduced, which
we compute using the probability density function~\eqref{chi2.proba} as
follows: $\delta^2=\int(\vx-\vx_0)^2p(\vx)\ud\vx$.

\begin{figure}
 \centering
 \includegraphics[width=0.7\linewidth]{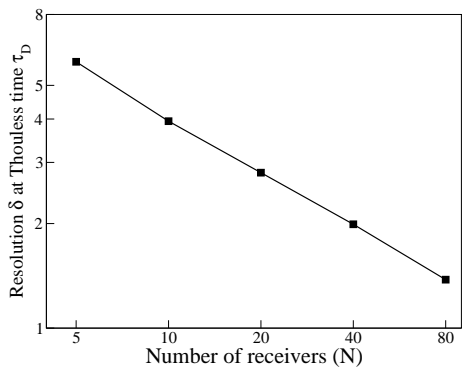}
 \caption{\label{f:accuracy-N} Spatial resolution at the coda time $t=\tau_D$,
 as a function of the number of receivers, where $\tau_D$ is the Thouless time.
 The typical setup for the numerical experiment is depicted in
 Figure~\ref{f:cercle}. The dots correspond to the values of the resolution
 for $N=5$, $10$, $20$, $40$ and $80$.
 The double logarithmic scale provides clear evidence
 of the relation $\delta\sim N^{-1/2}$.}
\end{figure}

In the vicinity of the change, we infer that the contributions of the terms
in~$e(\vx)$ are comparable and we deduce that $\delta\propto\epsilon$. Thus,
the precision with which the measurements are made directly influences the
precision with which the change is located. We will not study the dependence
of~$\delta$ with respect to~$\epsilon$ and we chose a value $\epsilon=0.01$
throughout the numerical study. Note that a uniform probability distribution
corresponds to a complete absence of information concerning the location of the
change, and gives the value~$\delta\simeq L$. The typical behaviour of the
resolution~$\delta$ as a function of the number of source-receiver pairs is
depicted in Figure~\ref{f:accuracy-N}. In the configuration described above,
each pair gives a comparable contribution to $e(\vx)$ so that $e(\vx)$ is
approximately proportional to $N$. Therefore in the ideal case described in our
example, we find that $\delta\propto N^{-1/2}$. 

Note that the resolution cannot be made arbitrarily small by increasing~$N$ at
will, because it is not possible to find an arbitrary number of source-receiver
pairs providing \emph{independent} data. The value~$N$ entering into the
scaling law $\delta\propto N^{-1/2}$ is the number of independent decorrelation
measurements. 

\subsection{Resolution versus coda time}
\begin{figure}
 \centering
 \includegraphics[width=0.9\linewidth]{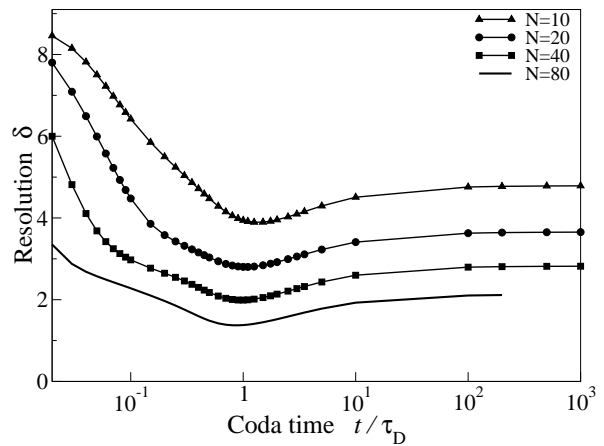}
 \caption{\label{f:accuracy-time}Spatial resolution obtained for
 the setup of Figure~\ref{f:cercle} for the values $N=5$, $10$, 
 $20$, $40$ and $80$ and for coda times varying from $2.10^{-2} \tau_D$ 
 to $10^3 \tau_D$. The time scale is logarithmic. A minimum
 of the resolution is found at $t\simeq \tau_D$. }
\end{figure}

The dependence of the resolution~$\delta$ with respect to coda time, is shown
on Figure~\ref{f:accuracy-time} for $\sigma=1$, $D=1$, $c=1$, $L=10$ and
$\epsilon=0.01$. The resolution exhibits a minimum at a time of order~$t_{\rm
min}=\tau_D$. For a given source-receiver pair, the coda time is the time that
has elapsed after the arrival of the ballistic wave. Shortly after the
ballistic arrival, the waves that reach the receiver have followed
``snake-like'' paths around the direct ray. In the early coda, the only
signals sensitive to the change are those for which the change is located along
the segment joining the source and the receiver. Later in the coda, the
diffuse waves arriving at the receiver have explored a larger volume of the
system. This qualitatively explains why~$\delta$ decreases with the coda time
$t$. At very late times, the formula \eqref{formula.diffusion} reveals that the
decorrelation for each source receiver pair saturates, as the exponential
factor tends to~$1$. The asymptotic spatial sensitivity to the change is
algebraic only. After reaching a minimum, $\delta$ increases because the
variations of~$\chi^2_n$ with respect to~$\vx$ decreases. The minimum
for~$\delta$ is found approximately at time~$\tau_D$, the Thouless time, after
which the whole system has been explored by the diffuse waves and yet~$Q$ still
exhibits large spatial variations. 

\subsection{Resolution versus cross-section}
\begin{figure}
 \centering
 \includegraphics[width=0.9\linewidth]{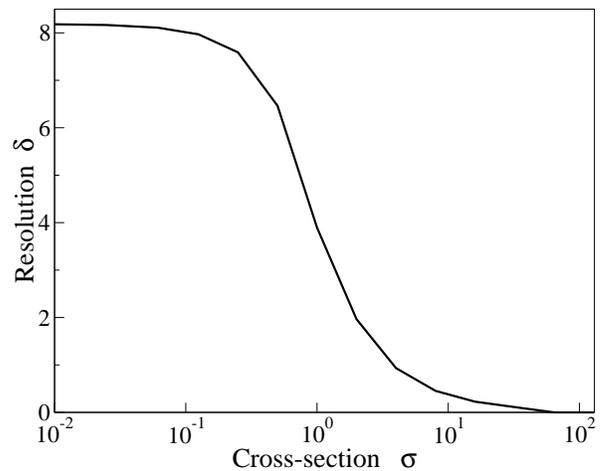}
 \caption{\label{f:accuracy-sigma} Optimal resolution~$\delta$
 as a function of the cross-section of the change~$\sigma$ obtained for the
 setup of Figure~\ref{f:cercle}. Other parameters
 of the simulations are $D=1$, $c=1$, $N=10$, $\epsilon=0.01$.}
\end{figure}

The scattering cross-section~$\sigma$ of the change also influences the
precision of the technique. We observe that the resolution~$\delta$ decreases
as~$\sigma$ increases. Note that when~$\sigma$ is very small, $\delta$ goes to
a value $\sim L$, meaning that it is not possible to detect the change. When
$\sigma\simeq L^2$, the cross-section is equivalent to the area of the system,
and locating a change has no physical significance in this limit. In
Figure~\ref{f:accuracy-sigma} we plot the variations of~$\delta$ at the optimal
time~$t=\tau_D$ as ~$\sigma$ varies from~$10^{-4}L^2$ to $L^2$. The other
parameters of the calculations are $D=1$, $c=1$, $\epsilon=0.01$, $N=10$. We
observe that the spatial resolution $\delta$ decreases by a factor 2 as the
cross-section increases from $10^{-2}$ to 1.

\subsection{Sensitivity to the value of the diffusivity $D$}
\label{s:diffusivity}
Our inversion technique depends heavily upon our ability to estimate the
diffusivity of the waves in the heterogeneous medium. Although the absorption
time $\tau$ does not enter into the final formula (\ref{formula.diffusion}),
let us remark that in practice $D$ and $\tau$ cannot be measured independently.
The diffusivity $D$ is the crucial physical parameter which enters into the
formula for the intensity propagator $P_d$ and controls the accuracy of the
energy propagation model of the medium. It is therefore important to quantify
the impact of errors in the diffusivity~$D$ on the accuracy of our method.
Even if we use an incorrect value for the diffusivity, our inversion procedure
still provides an answer for the position of the defect. The main issue is to
quantify to what extent the inferred position differs from the exact location
of the target. To address this point, we plot the spatial resolution and the
absolute error of the inversion for a wide range of values of~$D$ on a specific
example. 

\begin{figure}
 \centering
 \includegraphics[width=0.9\linewidth]{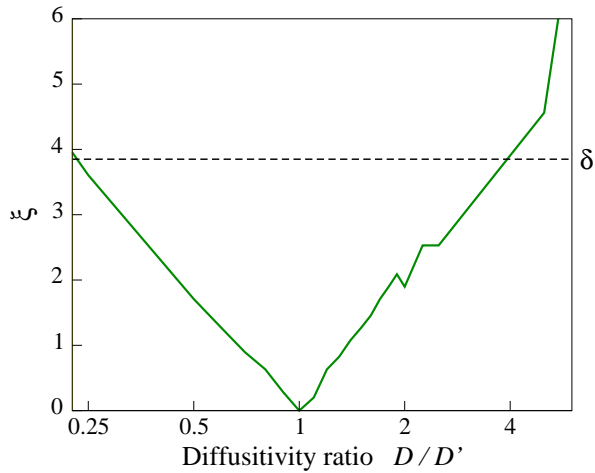}
 \caption{\label{f:accuracy-D} Effect of an estimation error on the 
 value of the diffusivity $D$ induced on the relocation of a target.
 Synthetic data were calculated with $D=1$ and inverted with modified 
 values of the diffusivity $D'$ ranging from 0.2 to 8. The other parameters
 of the simulation are $\sigma=1$, $L=10$, $N=10$, $\epsilon=0.01$ and the
 change is located at the position~$\vx_1$ (see Figure~\ref{f:cercle}). $\xi$
 is the distance between~$\vx_1$ and the point where~$\chi^2(\vx)$ is minimum.
 In the simulated configuration, the inversion technique remains accurate even
 if~$D'$ differs from~$D$ by a factor~$2$.} 
\end{figure}

We use the approach described in Section~\ref{s:resolution}. First, a
synthetic data set is computed with a value $D$ for the diffusivity. This
synthetic data set is then inverted for the location of the target using a
different diffusivity~$D'$. The change is located at the position~$\vx_1$
indicated on Figure~\ref{f:cercle}. The other physical parameters $L=10$,
$\sigma=1$, $N=10$, $\epsilon=0.01$ and $t$ have been adjusted to provide the
smallest spatial resolution~$\delta$. We call~$\xi$ the distance between the
change located by the inversion and $\delta$ is the resolution length. The
results of the simulation are displayed in Figure~\ref{f:accuracy-D}. It is
rather remarkable that an error on ~$D$ as large as a factor of~$2$ yields a
location of the change within one half of the resolution length. In this
specific but realistic example, the inversion technique is therefore very
robust against errors on the determination of~$D$. This constitutes a major
advantage of our method. Based on these results, we infer that spatial
variations of~$D$ within a factor of~$2$ will not affect the results
dramatically.

\section{Boundary conditions}
\label{s:boundaries}
The inversion technique presented in section \ref{s:inversion} relies on the
knowledge of the function~$P_d$, the diffusion kernel, which depends on the
boundary conditions of the system. For simplicity, we studied the LOCADIFF
technique in an infinite medium without taking into account the effect of
boundaries, which may not be realistic in applications.  An abundant literature
is dedicated to solving the diffusion equation in a wide range of
situations~\cite{crank}. In many cases of practical interest, sophisticated
techniques are required to provide an exact solution or a numerical
approximation up to a required accuracy. In the infinite medium, the
decorrelation~\eqref{def.Q} can be computed numerically. In the presence of
boundaries, it is more difficult to compute the Green's function because
translational invariance is lost. However, if the boundaries are flat, it
possible to construct the Green's function from the solution without boundaries
using symmetry arguments. In the general case, one has to solve the diffusion
equation for the geometry of the system, which is a problem for applied
mathematics in itself.

\begin{figure}[!hbtp]
 \centering
 \includegraphics[width=0.5\linewidth]{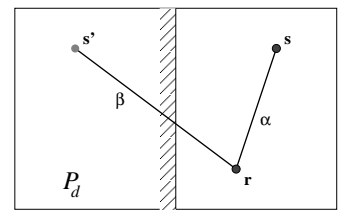}
 \caption{\label{f:boundary}
 Schematic representation of a boundary condition for~$P_d$. The image of
 the source~$\vs$ is noted~$\vs'$ and the arbitrary point is~$\vr$.
 The solid line is the solution in the infinite medium.}
\end{figure}

\begin{figure}[!hbtp]
 \centering
 \includegraphics[width=\linewidth]{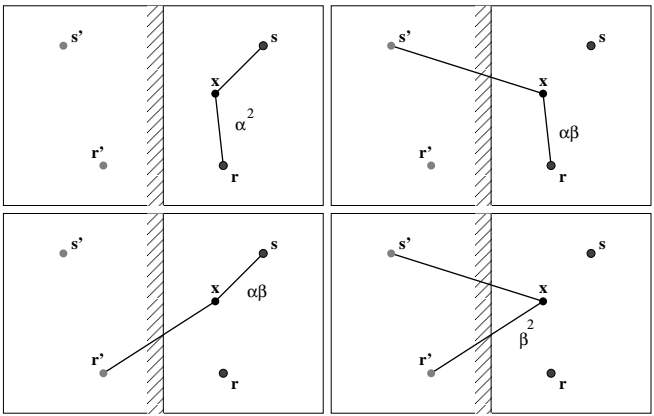}
 \caption{\label{f:boundaryQ}
 In presence of a single straight boundary, 
 the decorrelation function~\eqref{def.Q} involves four terms,
 coming from the product of two formula of the form~\eqref{e.solPBd}.
 We use the images of the source~$\vs$ and the receiver~$\vr$:
 Thanks to the symmetry of the diffusion equation, there is no
 need to introduce the image of the point $\vx$.}
\end{figure}

In the simple case of a single planar boundary, the solution~$P^B_d$ of the
diffusion equation of the semi-infinite medium, can be deduced from
$P^\infty_d$, using the technique of images: 
\begin{equation}
 P^B_d(\vs,\,\vr,\,t)=\alpha
 \left(P_d(\vs,\,\vr,\,t)+\beta P_d(\vs',\,\vr,\,t)\right)
 \label{e.solPBd}
\end{equation}
where $\vs'$ is the image of~$\vs$ with respect to the boundary (see
Figure~\ref{f:boundary}) and $\beta$ is a characteristic coefficient depending
on the nature of the boundary condition. For instance if the boundary is
absorbing, $\beta=-1$ and if it is fully reflecting, we have $\beta=1$. The
normalization coefficient~$\alpha$ is, in the case of constant diffusivity
\begin{equation}
 \alpha^{-1}=\frac{1+\beta}2+\frac{1-\beta}2
 \erf\left(\frac{d_{B,\vs}}{\sqrt{4Dt}}\right)
 \label{e.alpha}
\end{equation}
where $d_{B,\vs}$ is the distance from the source to the boundary.
Note that $\alpha$ is undetermined in the case where the conditions
 $\beta=-1$ and $d_{B,\vs}=0$ are met simultaneously.
The solution to the diffusion equation in presence of the boundary can be
plugged into the decorrelation expression~\eqref{def.Q} leading to four terms
(figure~\ref{f:boundaryQ}). Note that in case there are more boundaries, the
image technique requires to take into account infinitely many images. Other
techniques also lead to infinite series.

\section{Discussion}
\label{s:discussion}
In this section, we discuss issues related to the practical use of the LOCADIFF
technique as well as possible improvements. We first note that if the interval
between the records of $h_{ij}$ and $h'_{ij}$ is large, the medium may also
have experienced a global change, for instance a dilation due to a temperature
change. In this case, the computation of the decorrelation may be refined by
taking into account a global relative velocity change $\epsilon$, where
$\epsilon$ yields the maximum value of the correlation
\begin{equation}
 \frac{\int h_{ij}((1+\epsilon)u) \,h'_{ij}(u) \ud u}
 {\sqrt{\int h_{ij}(u)^2 \ud u
 \int h'_{ij}(u')^2 \ud{u'}}}.
\end{equation}
where the integrals are performed along the whole record \cite{larose2008}.

Another issue concerns the possible improvements on the inversion procedure.
Under the form presented in this article, the LOCADIFF technique only uses a
small time window in the signals. It would be of great interest to take into
account several time windows in the coda. This would provide more independent
data for the inversion procedure and may reduce the effect of noise. 

Finally, we point out that the kernel used in the inversion is computed from
the solution to the diffusion equation. In some simple geometries, like the
infinite medium, the solution is analytic and simple to compute. If the shape
of the medium is irregular, with possibly more complicated boundary conditions,
the kernels can only be approximated numerically. Alternatively, our approach
could benefit from recent developpements in the implementation of the radiative
transfer equation.

\section{Conclusion}
In this article, we have shown that it is possible to use the high sensitivity
of diffuse waves to detect, characterize and locate a small change in a
strongly scattering medium. Our technique uses the correlation of coda
waveforms recorded before and after the change. Based on a maximum-likelihood
approach, and a simple diffusion model, we demonstrate the possibility to
retrieve the position of the change along with its scattering cross-section.
We have also investigated the optimal values of the parameters that enter in
the inversion procedure, based on a simple setup where sources and receivers
are arranged on a circle surrounding the change. Three features have been
identified: 1) We found that the resolution scales with the inverse square root
of the number of sensors. 2) The technique provides the best results when the
correlation window is centered on the Thouless time of the system. 3) We
demonstrated that the technique is not very sensitive to errors in the
measurement of the diffusivity.

 Several aspects are still to be investigated. First, we have assumed that a
single change occurs in the medium, an assumption which is probably too
restrictive in some applications. In a straightforward generalization of our
technique to $n$ changes, the dimension of the parameter space scales like
$4n$ which in turn considerably increases the computation time. An alternative
route for the inversion has to be found. Second, we have made the assumption
of a point-like change. An extended change may not necessarily be equivalent to
a collection of point-like changes. Again, an alternative approach to the
inverse problem will be needed. We are currently investigating these two
issues.

Using 2D finite difference wave simulations, we have demonstrated that LOCADIFF
efficiently locates a small change in a multiple scattering environment. In a
seperate paper \cite{larose2010a}, experiments have also been conducted with
ultrasound in concrete. The change was a hole drilled in the sample, and the
LOCADIFF technique successfully retrieved its actual position. Other
applications in geophysics and material sciences can be envisaged.

\section*{Acknowledgments}
The authors thanks N. Tremblay and C. Sens-Sch\oe nfelder for discussions. This work was supported by the ANR JC08\_313906 \textit{SISDIF} grant.

\appendix
\section{Derivation using Bayesian inversion}
We shortly derive here the density of probability density~\eqref{chi2.proba}
using a Bayesian inference. In this calculation, we suppose that there is a
change at an unknown position~$\vx$. The values of the measurements~$K^m_{ij}$
are accurate up to an error order~$\epsilon$ such that they are distributed
around the numerical value $K_{ij}(\vx,t)$ according to a standard error
function. 

\begin{equation}
 p(K^m_{ij} | \vx)=\frac1{\sqrt{2\pi}\epsilon}
 \exp\left[-\frac{(K_{ij}(\vx,\,t)-K^m_{ij})^2}{2\epsilon^2}\right].
 \label{e.pKx}
\end{equation}
Each pair $(i,j)$ provides an independent information. The Bayesian inversion
consists in finding the probability density of~$\vx$ knowing the values
of~$K^m_{ij}$, namely to compute $p(\vx | \{K^m_{ij}\})$. Let us call
$p_n(\vx)$ the probability density for the position of the change when~$n$
source-receiver pairs have been taken into account. Before measurement, the
probability of the location of the change is uniform in the whole medium, so we
have~$p_0(\vx)=\frac1V$ ($V$ is the volume). Suppose we know $p_{n-1}(\vx)$ and
let us compute the joint probability of~$\vx$ and $K_n$ using Bayes' formula.
We use the two relations: 
\begin{align} 
 p(\vx,K^m_n|K^m_1,\dots\,K^m_{n-1}) & = 
 p_{n}(\vx)p(K^m_{n}) \label{e.Bayes1}\\
 p(\vx,K^m_n|K^m_1,\dots\,K^m_{n-1}) & = 
 p(K^m_n|\vx)p_{n-1}(\vx) \label{e.Bayes2} 
\end{align}
Integrating~\eqref{e.Bayes1} over~$\vx$ we can compute $p(K^m_{n})$ as
\begin{equation}
 \left(\int_V p_{n}(\vx) \ud\vx\right) p(K^m_{n})=
 \int_V p(\vx,\,K^m_{n}|K^m_1,\,\dots K^m_{n-1}) \ud \vx
 \label{K11}
\end{equation}
The integral of $p_{n}(\vx)$ is equal to~$1$ so we conclude that,
using~\eqref{e.Bayes2},
\begin{equation}
 p_{n}(\vx)=\frac{p(K^m_{n}|\vx)p_{n-1}(\vx)}
 {\int_Vp(K^m_{n}|\vx)p_{n-1}(\vx)\ud\vx}.
 \label{e:inversionBayes}
\end{equation}
Therefore we have a recurrence scheme yielding the distribution of
probability~$p_N(\vx)$:
\begin{equation}
 p_N(\vx)=\frac{\prod_{n=1}^N p(K^m_n|\vx)}
 {\int\prod_{n=1}^N p(K^m_n|\vx)\ud\vx}
\end{equation}
which gives Equation~\eqref{chi2.proba} after replacing the probabilities with
expression~\eqref{e.pKx}. 

\bibliography{locadiff}

\end{document}